# Technology Parks' Potential for Small and Medium Enterprises

Anna V. Vilisova
weianna@yandex.ru

Fu Qiang
fuqiang@cqu.edu.cn

**Abstract.** Being one of the most important factors of country's economic growth, innovations became one of the key vectors in Russian economic policy. In this field technology parks are one of the most effective instruments which can provide growth of innovative activity in sectors, regions and economies.

In this paper, we made a model that allows us to evaluate the effect of technology parks in the country's economy and it's potential for small and medium enterprises. The model is based on a system of coupled equations, whose parameters are estimated on the statistical data that reflect the activity of the economic entity, in an environment of this entity the technology parks are acting. Typically, there are regression equations linking a number of economic factors with some output indicators. We analyzed the property of increasing the share of surviving small and medium enterprises for Russian conditions as one of the effect of technology parks and built a working model for estimating the maximum (limit) values of the effect.

**Introduction.** Intensive development of the economies of different countries makes it necessary to maintain their competitive advantage at a high level, which is quite difficult to do without the use of innovations.

In many countries, there are special areas (technology parks - TP, science parks, etc.), providing the necessary conditions to stakeholders (start-ups, small businesses, private entrepreneurs) to efficient creation of innovative solutions in various sectors of the economy [1].

In Russia the attempts to create technology parks started from the "bottom" in the late 90's, but legal support in the form of legislation appeared only in 2005-2006 [2]. There was created an informal social organization called Association of Russian industrial parks, designed to provide support for their creation, existence and development. However, up to date there is no effective unified targeted mechanism to support the system of parks' functioning, aimed at the final result - growth of the country's economy [3, 4].

Investments in TP, as in the infrastructure of the innovation industry, in many countries implemented by the government, and in some cases - at the expense of private capital [5].

Creation of TP is often a venture project, because almost always it is difficult to estimate whether the investments will lead to creation of innovative products. In some cases, before starting the project investors carried out preliminary estimates of the expected economic impact of a TP, but in many cases, these calculations are not carried out due to lack of appropriate methodological tools [6].

In this paper we attempt to build a model that evaluates the effect of technology parks in the country's economy. However, the proposed model also can be used for smaller-scale economic units - regions, cities, etc. The model is based on a system of coupled equations, whose parameters are estimated on the statistical data reflecting the activity of the economic subject, in an environment of TP. Typically, the mentioned equations are regression dependences, connecting a number of economic factors, with some output indicators.

**The concept of the model building.** Considering that the TP are the elements of the country's economic system, we will consider their private data as input variables affecting the integrated, more general indicators of the economy (output variables).

The model reflects the links, which are typical for the economy of a country. These links are correlational in nature and can only approximate the real-world complicated mechanisms of the economic environment as a complex organizational system. The formal purpose of our study is to build a model of the influence of the characteristics of small and medium enterprises (SMEs[1]) operating in the TP, on the country's GDP. First, consider a simple characteristic, i.e. the total number of the small businesses. In the future, the same technology can be used to study the influence of other parameters of SMEs on output indicators of the economy.

---
[1] In this paper terms "small and medium enterprises" (SME) and "small-scale enterprises" (SSE) considered to be the same.

The models will be constructed in the form of a regression represented by polynomials of the first and sometimes the second degree and linear in the parameters [7, 8].

Evaluation will be based on the correlations (and the corresponding regression) between the factors of TP and the total economic indicators (such as GDP). We will represent their relationships in the form of a block diagram shown in Figure 1.

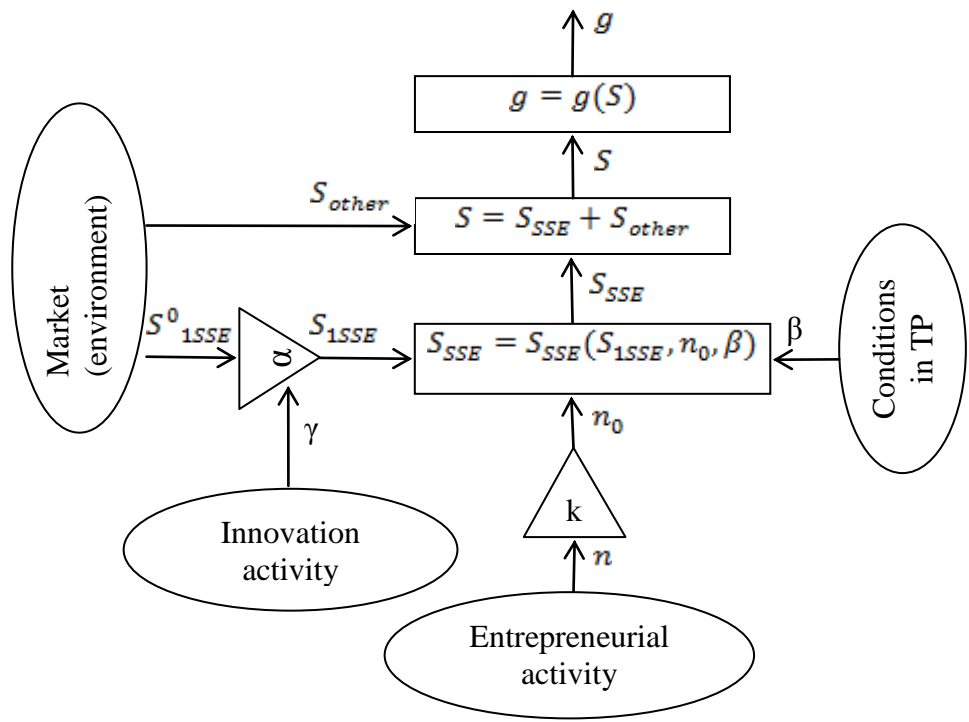

Fig. 1. Interconnection of elements in the output indicator's evaluation circuit

Let's explain the logic and the basic elements of the scheme. Entrepreneurial activity leads to an amount of SME start-ups ($n$), of which only $n_0$ survives, which is a part of the original amount $k$ ($k \in [0,1]$). The turnover of the whole set of SME ($S_{SSE}$) depends on the average turnover (annualized) of one SME ($S_{1SSE}$) and also depends on the number of "SME-survivors". In addition, the amount $S_{SSE}$ is affected by availability of technical parks in the country. The degree of influence depends on many factors, the net effect of which is presented by the coefficient [β], which can raise the rate of survived SMEs till one ($\beta \in [0, (1-k)]$).

GDP represented by the value $g$, which depends on the turnover of all enterprises $S$, $g = g(S)$. Turnover $S$ consists of the turnover of all SMEs ($S_{SSE}$) and the turnover of all other companies $S_{other}$.

Turnover of each company, along with many other factors, depends on the value innovative activity ([γ]) of the economic subject, i.e. in the country, industry, region, etc. The innovation activity of SME leads to increase in [α] times of the average turnover of SME ($S^0{}_{1SSE}$), which in turn leads to GDP increasing.

In the scheme of the variety of factors of GDP increasing, it puts more emphasis on those (two), which are caused by the activity of TP and the conditions created there (in TP) for the SME start-ups. The TP's activity increases the proportion of SME survivors ([β]) and also increases the realization of their innovative potential, so that leads to increasing in the average turnover of one SME in [α] times.

Thus, by the model of estimating the technology parks' potential (METPP) we mean a set of interrelated individual models represented by the individual equations, these equations establish functional relationships between the set of parameters of the economic subject and its indicators. The part of the parameters related to the technology parks, and that allows making estimates of the degree of their influence on the important measures of the economic subject. The limit of the improvements of these indicators by varying the parameters related to the TP, and represents the potential due to factors of TP.

We assume that all of the dependencies and the required values of the coefficients can be defined as a statistical evaluation and /or regression equations that reflect the correlation of the model's parameters. To obtain the necessary assessments we should use the official data published by the statistical services of the economic entities. For Russia it's Russian Federal State Statistics Service [9].

**The indicators of the economic entities' effectiveness.** As far as most of the economic entities are complex systems [10], the quality of their functioning is usually described with not only one factor, but with a combination of all factors. However, for ease of analysis specialists often choose one the most important indicator. Among those in any country there is such an integral indicator such as gross domestic product (GDP), which we will consider as an output indicator of the economic subject, the country. Usually in published statistical data there is nominal GDP (in prices of this year). However, in order to compare the results of several years the inflation factor (increase in prices from the previous year) is usually taken into account. On the base of it the deflator is being counted, it is the coefficient reflecting the change in prices relative to a fixed year (e.g., 2013). Then the real GDP will reflect the amount, adjusted by the amount of the deflator (see the approximate form in Table 1).

Table 1

A part of Russian GDP data for a number of years (in billions of rubles).

| Year | Nominal GDP | Inflation | Total inflation (by 2013) | Deflator (by 2013) | Real GDP (by 2013) |
|---|---|---|---|---|---|
| 2010 | 45173 | 1.085 | 1.223 | 0.817 | 55278 |
| 2011 | 54586 | 1.060 | 1.128 | 0.887 | 61564 |
| 2012 | 56769 | 1.064 | 1.064 | 0.940 | 60403 |
| 2013 | | 1 | 1 | 1 | |

Changing in the real GDP ($g$) by the years ($t$) can be obtained by regression analysis [7]. Thus, the regression analysis typically has quite a linear approximation:

$$g = a_0^g + a_1^g t. \quad (1)$$

Here, the index of coefficients in the regression equation reflects the number of the parameter (or variable) in the regression equation. We will call this dependency the Model of the Indicator's Dynamics (MID), and will write generally as $g(\bar{a}, t)$, where $\bar{a}$ is vector (set) of the model's coefficients. In this paper, an indicator will be referring to the country's GDP.

The **purpose of the work** is to build a dependence of the index on the time and number of factors, in particular, the presence of TP, as an environment, providing conditions for the formation of SME and other startups.

The **problems** solved in the work to achieve the goal are to build local models, which allow us to set dependencies of the auxiliary parameters and indicators on the base of the available statistical data (for example, for Russia it's Federal State

Statistics Service [9]), so that will give the opportunity to build a Parametric Model of the Indicators' Dynamics (PMID) similar to (1).

**Local models of industrial development and the development of entrepreneurial activity.** The main contribution to GDP is made by industrial enterprises, and small businesses (including innovative businesses and residents of TP) are the part of them. Therefore, the parameters of the technology parks will affect indirectly (through the properties of their residents, as a rule, of the small innovative enterprises) the output parameters of the economic subject [1]. For this study the interesting indicators are: number of enterprises and turnover.

Using the statistical data and regression analysis we can obtain the dependence of the changings of the number ($n$) of small enterprises by years ($t$), we will call it the Small and medium Enterprises' Generation Model (SMEGM). Generally it can be written in the form of $n(\bar{c},t)$, where $\bar{c}$ is vector of the model's parameters. Practice shows that this dependence can be well approximated by a second-order polynomial regression:

$$n = c_0 + c_1 t + c_2 t^2. \quad (2)$$

In the economies of many countries, the overall increase in the number of enterprises is mainly provided by small private companies. But different countries have their features. For example, small private enterprises in Russia in 90's did not make any significant contribution to the overall turnover and, respectively, in the country's GDP [1]. The turnover was mainly provided by the large state-owned or municipal enterprises.

In addition to the number of enterprises, an important parameter that affects the output indicators, is the actual turnover of enterprises of all patterns of ownership ($S$) considering the deflator. According to statistics from the state statistical service (for example, in Russia – Rosstat, Russian Federal State Statistics Service [9]) we can construct a regression model in which the input (independent) variable is the turnover of the organizations ($S$), and the output (dependent) variable is GDP ($g$). A linear regression model (we will call it the Flip Model of Index, FMI) generally can be written as $g(\bar{b},S)$, in particular linear form it would be:

$$g = b_0 + b_1 S. \quad (3)$$

Since one of the main functions of TP is to support the process of formation of small enterprises [11, 12], so in the considered chain of local models the important part is the regression dependence of turnover of small businesses on the whole amount of enterprises. Based on the logic of relations between all and small businesses, this local model (we will call it the SME's Flip Mode, SMEFM) should look like a simple proportional relation in a form of:

$$S_{SSE} = d\,S. \quad (4)$$

Here, $d$ is the proportion of SME in the total number of enterprises of the economic subject.

**Parameters of small businesses.** In many countries, SMEs may be represented by several specific groups, different in the number of parameters. For example, in Russia SMEs are usually divided into the following categories:

- individual entrepreneurs (the number of employees is up to 5 people);
- micro-enterprises (number of employees up to 16 people and revenues are up to 60 million rubles.);
- small businesses (the number of 16 to 100 people, or the proceeds from 60 to 400 million rubles.);
- medium-sized enterprises (the number from 100 to 500 people or turnover more than 400 million rubles.).

An important characteristic of the SME is the average amount of revenue (AAR) of one enterprise ($S_{1SSE}$).

It should be noted that the vast majority of resident enterprises in TP usually is not among the medium-sized enterprises, which means that medium-sized enterprises should be excluded from the AAR evaluation.

**The Potential of TP**

The main effect of technology parks supporting small businesses should be expressed in increasing the share of the survived businesses after the passage of the first stages of their formation and development. Therefore, if we know the number of active SME ($n_0$), which arose in the almost complete absence of TPs, then we can estimate their limiting number ($n_0^{TP}$) for the case when all of them were formed in the environment of TPs:

$$n_0^{TP} = n_0 \frac{k+\beta}{k}, (5)$$

where $k$ is the proportion of SMEs who survived in the "wild" environment (without technology parks), and [β] is the proportion of SMEs who survived due to the factor of the TPs presence in the economic entity (for example, in the country). This dependence will be called SME's Potential Number Model (SMEPNM). It should be noted that the equation (5) is of a general nature so variant $\beta = 0$ corresponds to the absence of TP.

Counting the effect of increased share of the survived start-ups in the TPs allows us (taking into consideration also the local models written before) to estimate the contribution of technology parks in the total indicator (e.g., the country's GDP). We will show how this can be done by making PMID.

The regression dependence of the final index on enterprises' turnover (FMI) can be represented by the expression (3), which does not contain a time parameter ($t$) in the explicit form, i.e. is not formally dynamic but static. However, the variable turnover ($S$), being a part of the expression (3), depends on time and applies to all businesses, including the SME. To isolate the effect of TP only for SME, we will divide $S$ into two components - the turnover $S_{SSE}$ for SME only (see (4)) and the turnover $S_{other}$ for other businesses. $S_{other}$ will be defined from (4) as:

$$S_{other} = (1-d) S. (6)$$

The proportion of the turnover of all the SME in (3) we will express using the average turnover of one SME $S_{1SSE}$ and the number of them $n_0^{SSE}$ (see (5)):

$$S_{SSE} = S_{1SSE} n_0 \frac{k+\beta}{k}. (7)$$

Substituting (6) and (7) into (3) and adding the time parameter ($t$), to reflect the change the turnover over the time, we obtain an expression for the value of the indicator at an arbitrary year $t$:

$$g(\beta,t) = b_0 + b_1 \left((1-d)S(t) + S_{1SSE}(t)n_0 \frac{k+\beta}{k}\right). (8)$$

The obtained expression is structurally similar to (2), but includes a parametric connection of the output indicator (GDP) with the characteristics of SME and TP.

From the expression (8) we see that the increment of the output index with the parameter [beta] changing, where [beta] reflects the influence of the TP environment, is linear and can be represented by a simple regression equation:

$$\Delta g(\beta,t) = g(\beta,t) - g(0,t) = b_1 S_{1SSE}(t) n_0 \frac{1}{k} \beta = \varepsilon \beta. \quad (9)$$

In general, the model (9) connects the output index's increment and several input parameters, so the parametric form of the potential's model can be written as follows:

$$\Delta g(b_1, S_{1SSE}, n_0, k, \beta, t) = b_1 S_{1SSE} n_0 \frac{1}{k} \beta. \quad (10)$$

Then, for any occasion the potential can be calculated by forward substituting the values of all the parameters into (10). Thus, each parameter can be time-dependent. Since the parameters of the model in the actual economic practice are calculated according to statistical data, as a rule, by means of regression analysis (see local models (1) - (4)), so for computational convenience, we will present a simplified logical chain of these calculations.

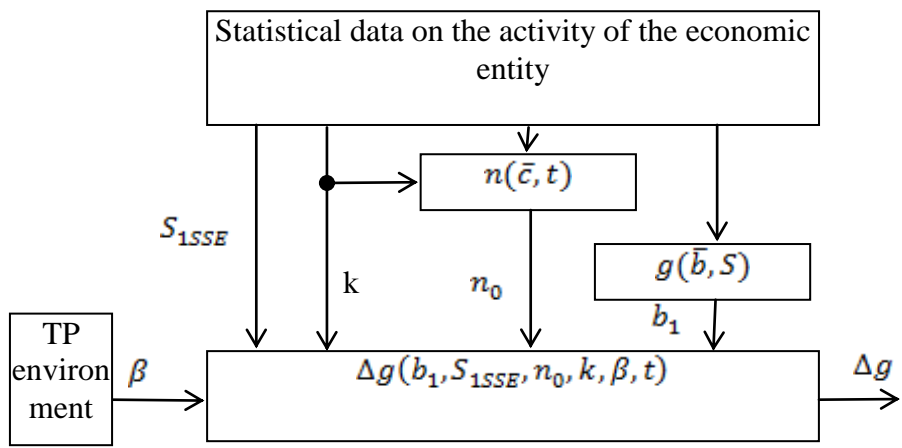

Fig. 2. The relationship of local models in the scheme of computation of TP potential

**Conclusion**

1. Being the property of the TP potential, the creation of conditions for the survival of start-ups (SME) by the presence of business incubators, preferences and benefits for the residents of TP, etc. in their structure, increases the proportion of survivors SME. The property of increasing the share of surviving SME, as one of the effects of TP, has been analyzed for Russian conditions and the estimates of the maximum possible values have been got. This type of TP effect is treated as if it came instantly (in "what if ..." mode). This approach allowed us to build a working model for estimating the maximum (limit) values of the effect.

2. The paper contains a sequence of local models that can be built according to the statistical reporting of economic activity of the analyzed subject.